\documentclass[bib]{statapress}

\usepackage[crop,newcenter,frame]{pagedims}
\usepackage{enumitem}
\setlist[enumerate]{nosep, itemsep=0pt, parsep=0pt, topsep=0pt}
\setlist[itemize]{nosep, itemsep=0pt, parsep=0pt, topsep=0pt}
\usepackage{sj}
\usepackage{amsfonts}

\usepackage{epsfig}
\usepackage{subcaption}

\usepackage{stata}

\usepackage{amsmath}

\usepackage{shadow}

\sjsetissue{$vv$}{$ii$}{$mm$}{$yyyy$}

\usepackage{threeparttable}
\usepackage{tabularx}

\usepackage{xcolor}

\begin{document}

\inserttype[st0001]{article}

\author{F. Gunsilius and D. Van Dijcke}{
  Florian Gunsilius \\ Emory University \\ Atlanta, Georgia/US \\ fgunsil@emory.edu
  \and
  David Van Dijcke \\ University of Virginia \\ Charlottesville, Virginia/US \\ eju5az@virginia.edu
}

\title[disco]{disco: Distributional Synthetic Controls}

\maketitle

\begin{abstract}
The method of synthetic controls is widely used for evaluating causal effects of policy changes in settings with observational data. Often, researchers aim to estimate the causal impact of policy interventions on a treated unit at an aggregate level while also possessing data at a finer granularity. In this article, we introduce the new \texttt{disco} command, which implements the Distributional Synthetic Controls method introduced in \citet[\textit{Econometrica} 91: 1105--1117]{gunsilius2023distributional}. This command allows researchers to construct entire synthetic distributions for the treated unit based on an optimally weighted average of the distributions of the control units. Several aggregation schemes are provided to facilitate clear reporting of the distributional effects of the treatment. The package offers both quantile-based and cumulative distribution function--based approaches, comprehensive inference procedures via bootstrap and permutation methods, and visualization capabilities. We empirically illustrate the use of the package by replicating the results in \citet[\textit{Review of Economics and Statistics}, forthcoming]{van2024return}.

\keywords{st0001, disco, disco\_estat, disco\_plot, disco\_weight, synthetic control, treatment effect, distributional analysis, causal inference, optimal transport}
\end{abstract}

\section{Introduction}
Synthetic control methods, pioneered by \citet{abadie2003economic} and formalized in \citet{abadie2010synthetic}, have revolutionized policy evaluation with aggregate data. These methods have become a cornerstone of modern causal inference, particularly in settings where traditional regression approaches are infeasible due to a limited number of treated units or the lack of a good counterfactual control group. While these methods have proven invaluable for estimating average treatment effects, they traditionally focus on matching means, potentially overlooking rich heterogeneous effects across the distribution of outcomes that are often of primary interest to researchers and policymakers.

The Distributional Synthetic Controls (DiSCo) method \citep{gunsilius2023distributional} addresses this limitation by extending the synthetic control framework to match entire outcome distributions rather than just means. This extension is particularly valuable in empirical settings where researchers have access to repeated cross-sectional data within aggregate units but cannot track individual units over time. Such scenarios are common in policy evaluation, where interventions occur at an aggregate level (e.g., states, counties, or firms) but researchers have access to individual-level or more granular data within these units. In these contexts, treatment effects may vary substantially across the distribution of outcomes, making that distribution a crucial object of study.

This article introduces the \texttt{disco} command in Stata, which implements the Distributional Synthetic Controls method. We first provide a practical overview of the underlying methodology, then lay out the command's syntax and usage, and illustrate its use in an empirical application. A companion R package is available at \citet{discosR}.

\section{Methodological Background} \label{sec:methodology}

\subsection{Overview and Notation}

Consider a panel of $J+1$ units (for example, firms) observed over $T$ periods. One unit (conventionally labeled as unit 1) receives treatment after period $T_0$ (or equivalently in period $t_0 = T_0+1$), while the remaining $J$ units serve as potential controls. For each unit $j$ and time period $t$, we observe individual-level outcomes $Y_{ijt}$ (for example, employee wages). From these, we can construct the cumulative distribution function (CDF) $F_{Y_{jt}}$
\[
F_{Y_{jt}}(y) = P(Y_{jt} \leq y)
\]
which gives the probability $P$ that the outcome of unit $j$ at time $t$, $Y_{jt}$, is less than or equal to $y$, and its corresponding quantile function $F^{-1}_{Y_{jt}}$,
\begin{equation} \label{eq:pseudo_inverse}
F^{-1}_{Y_{jt}}(q) = \inf\{x \in \mathbb{R}: q \leq F(x)\}
\end{equation}
which captures the value in the support of $Y_{jt}$ that is larger than $q \times 100$ percent of the distribution.  
These functions capture the entire distribution of outcomes within each aggregate unit, preserving information about heterogeneity that would be lost by focusing only on means.

The goal is to estimate what the distribution of outcomes in the treated unit would have been in the absence of treatment. We denote this counterfactual distribution's CDF as $F_{Y_{1t,N}}$ and its quantile function as $F^{-1}_{Y_{1t,N}}$ for periods $t > T_0$. The DiSCo method constructs entire synthetic distributions for the treated unit using optimally chosen weights on the control distributions.

\subsection{Quantile-Based Approach}

The quantile-based approach constructs the counterfactual by finding optimal weights $\lambda_j$ that create a synthetic control distribution via a weighted average of control unit quantile functions:
\begin{equation} \label{eq:synthetic_quantile}
F^{-1}_{Y_{1t,N}}(q) \;=\; \sum_{j=2}^{J+1} \lambda^*_j F^{-1}_{Y_{jt}}(q) 
\quad \text{for } q \in (0,1)
\end{equation}
These weights are chosen to minimize the squared distance between the treated unit's pre-treatment quantile function and the weighted combination of control unit quantile functions, corresponding to minimizing the 2-Wasserstein distance,
\begin{equation} \label{eq:2-wasserstein}
\boldsymbol{\lambda}^*_t 
\;=\; 
\operatorname{argmin}_{\lambda \in \Delta^J} 
\int_{\mathrm{qmin}}^{\mathrm{qmax}} 
\Bigl|\sum_{j=2}^{J+1} \lambda_j F^{-1}_{Y_{jt}}(q) \;-\; F^{-1}_{Y_{1t}}(q)\Bigr| 
\, dq
\end{equation}
for $t \leq T_0$, where $(\boldsymbol{\lambda}^*_t) = (\lambda^*_{1t}, \ldots, \lambda^*_{jt})$ is the vector of $J$ weights (one for each control unit), and $\Delta^J$ is the $J$-dimensional unit simplex, which constrains the weights to satisfy $\sum_{j=2}^{J+1} \lambda_j =1$ and $\lambda_j \geq 0$ for all $j>1$. Then, the overall synthetic control weights are computed as a simple average over the pre-treatment weights,
\[
\lambda^*_j = \frac{1}{T_0} \sum_{t=1}^{T_0} \lambda^*_{jt}
\]
Here, $\mathrm{qmin}$ and $\mathrm{qmax}$ default to 0 and 1, respectively, matching the entire distribution. In practice, researchers may sometimes wish to match or conduct inference on specific parts of the distribution. To accommodate this, the command allows users to specify the \texttt{qmin} and \texttt{qmax} options, restricting the integral to an interval $[a,b] \subset [0,1]$.

To relax the positivity constraint and allow for extrapolation as in \citet{doudchenko2016balancing}, the user can specify the \texttt{nosimplex} option in the main command.   
The primary motivation for choosing this particular objective function is that it has the intuitive interpretation of a ``regression of quantile functions.'' Alternatively, one can interpret the problem as finding the set of weights that makes the treated unit's quantile function as similar as possible to the weighted average of the control units’ quantile functions \citep[p. 1109]{gunsilius2023distributional}. This aligns with the interpretation of the weighted average quantile function, $\sum_{j=2}^{J+1} \lambda^*_j F^{-1}_{Y_{jt}}(q)$, as a Wasserstein barycenter \citep{agueh2011barycenters}. 

Note that problem \eqref{eq:2-wasserstein} is ``distributional'' in the sense that it assigns a single weight to each quantile function in its entirety, which can be seen from the outer integration over $y$. Thus, the synthetic control is truly a weighted average of distributions, rather than a pointwise average of quantiles as in \citep{chen2020distributional}. The benefits of working with full distributions as the fundamental objects of estimation are: (1) it corresponds to the classical synthetic controls approach of assigning one single weight to every unit $j$; and (2) it allows for individuals to drop in and out of a unit, meaning one need not track individuals over time. This accommodates settings with entry and exit (e.g., employees leaving firms, as in \citet{van2024return}), and settings where the researcher may only have a repeated cross-section rather than a balanced panel of observations. 

In practice, the integral in \eqref{eq:2-wasserstein} must be simulated, resulting in the empirical problem,
\begin{equation} \label{eq:integral_simulation}
\boldsymbol{\lambda}_i^*=\underset{\bar{\lambda} \in \Delta^J}{\operatorname{argmin}} \frac{1}{G} \sum_{g=1}^G \sum_{j=2}^{J+1} \left| \lambda_{j t} F_{Y_{j_t}^{-1}}^{-1}\left(V_g\right)-F_{Y_{1t}}^{-1}\left(V_g\right)\right|^2
\end{equation}
where $V_g$ are $G$ independent draws from the uniform distribution on $[0,1]$, with $G$ large. In practice, the command simply uses uniformly spaced draws from $[0,1]$. We also use a single parameter $G$ as both the number of simulation draws for the integral and the number of grid points on which we evaluate the synthetic control distributions in practice. 

\subsection{CDF-Based Approach}

As an alternative, one can match the cumulative distribution functions directly, minimizing the 1-Wasserstein distance:
\begin{equation} \label{eq:1-wasserstein}
\boldsymbol{\lambda}^*_t 
\;=\; 
\operatorname{argmin}_{\lambda \in \Delta^J} 
\int_{-\infty}^{\infty} 
\left|\sum_{j=2}^{J+1} \lambda_j F_{Y_{jt}}(y) \;-\; F_{Y_{1t}}(y)\right| 
\, dy
\end{equation}
The synthetic control distribution function is then estimated analogously to \eqref{eq:synthetic_quantile} as
\[
F_{Y_{1t,N}}(y) \;=\; \sum_{j=2}^{J+1} \lambda^*_j F_{Y_{jt}}(y) 
\quad \text{for } y \in \mathrm{supp}(Y)
\]
with $\mathrm{supp}(Y)$ the support of the outcome variable $Y$. Of course, one can easily obtain a synthetic control quantile function from this by using the pseudo-inverse in \eqref{eq:pseudo_inverse}, and vice versa for \eqref{eq:synthetic_quantile}. Indeed, the command always computes both synthetic quantile functions and CDFs, regardless of whether the \texttt{mixture} option is specified, and stores them in \texttt{e(quantile\_synth)} and \texttt{e(cdf\_synth)}.

\begin{figure}[ht!]
\centering
\begin{subfigure}[b]{0.6\textwidth}
\includegraphics[width=\textwidth]{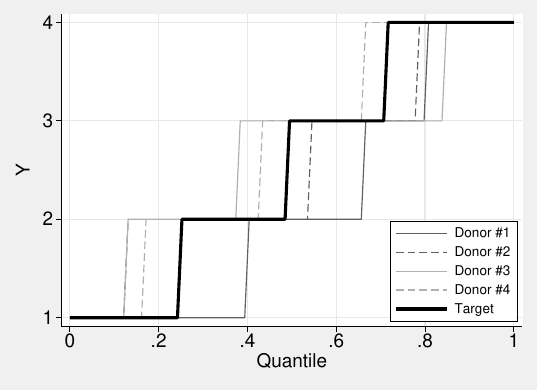}
\caption{Target and Control Quantile Functions}
\label{fig:mixture_raw_quantiles}
\end{subfigure} \\
\begin{subfigure}[b]{0.49\textwidth}
\includegraphics[width=\textwidth]{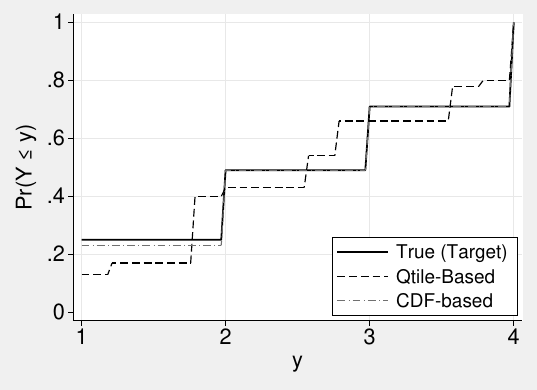}
\caption{Target and Synthetic CDFs}
\label{fig:mixture_cdf}
\end{subfigure}
\begin{subfigure}[b]{0.49\textwidth}
\includegraphics[width=\textwidth]{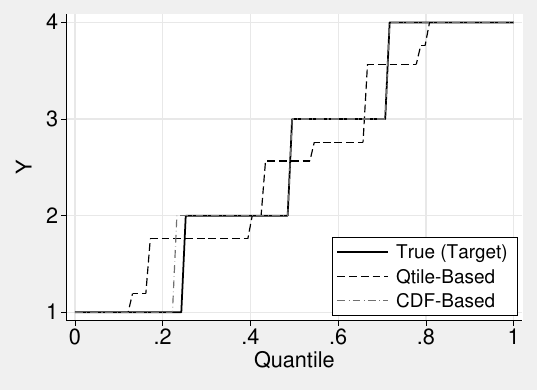}
\caption{Target and Synthetic Quantile Functions}
\label{fig:mixture_quantile}
\end{subfigure}
\caption{Illustrative Simulation of \texttt{mixture} Option}
\end{figure}

The CDF-based approach is often advantageous for discrete outcomes or settings where preserving specific portions of the distribution is critical \citep[\S 4.3]{gunsilius2023distributional}. To have the command solve \eqref{eq:1-wasserstein} instead of \eqref{eq:2-wasserstein}, one can specify the \texttt{mixture} option. A prominent example where the mixture (CDF-based) approach is useful is when the outcome is a categorical variable. In that case, a weighted average of quantile functions, as computed in \eqref{eq:2-wasserstein}, will interpolate between the support points, which is undesirable. A “mixture” of distribution functions, however, restricts the weighted average to the same support points as the “donor” distribution functions (see Figure 1 in \citet{gunsilius2023distributional}). We further illustrate this with a simple simulation in Stata: our target (treated) distribution is a discrete uniform distribution with equal probability mass on $\{1,2,3,4\}$, i.e.,
\[
Pr(Y_t=i) = \frac{1}{4} \text{ for } i=1,\ldots,4
\]
Our donor distributions take this target distribution but place more mass at one of the support points each, i.e.,
\[
Pr(Y_j = i) = 
\begin{cases}
    1/5 & \text{if } i \neq j, \\
    2/5 & \text{if } i = j, 
\end{cases} 
\]
for $i,j=1,\ldots,4$. The “raw” control and target quantile functions corresponding to these random variables are plotted together in Figure \ref{fig:mixture_raw_quantiles}. The synthetic CDFs and quantile functions, constructed by either solving \eqref{eq:2-wasserstein} (dashed) or \eqref{eq:1-wasserstein} (dash-dot), are plotted against the target distribution (solid) in panels \ref{fig:mixture_cdf} and \ref{fig:mixture_quantile}. As clearly shown, the CDF-based synthetic controls correctly preserve the support of the control (donor) functions, while the quantile-based approach undesirably interpolates between support points. Of course, such interpolation \textit{would} be desirable for a continuous outcome variable. 

Finally, note that when the outcome variable is categorical, one should specify the \texttt{g} and \texttt{m} options to reflect the number of support points. For instance, if the variable takes values on the integers from 1 to 10, one would specify \texttt{g(10)} and \texttt{m(10)}. This ensures that the integral in \eqref{eq:1-wasserstein} correctly sums over the support points. The code expects evenly spaced support points for this to work. If the outcome variable is not evenly spaced, one can normalize it without affecting the estimates. Furthermore, when creating graphs for categorical variables with \texttt{disco\_plot}, one may want to specify the \texttt{categorical} option to change the graph type to a bar plot instead of a line plot. 

\subsection{Inference}

\subsubsection{Permutation Test}

When the \texttt{permutation} option is specified, the command carries out the distributional version of the classical permutation test for synthetic controls proposed in \citet{gunsilius2023distributional} and further developed in \citet{van2024return}. The idea is intuitive: recompute the DiSCo estimator anew for each of the $J$ control units, while pretending that the given control unit was, in fact, treated. If the treatment effect on the treated unit is not driven by statistical noise, the estimated treatment effect on the true treated unit should be an extreme value of the resulting distribution of “placebo” treatment effects. 

The only complication in the distributional setting is that each single permutation delivers an entire distribution of treatment effects, so we need a way to aggregate these into a single statistic that we can rank. Echoing the use of the root mean squared prediction error in \citet{abadie2010synthetic}, \citet{gunsilius2023distributional} proposed using the $2$-Wasserstein distance. Then, an analogous permutation test to \citet{abadie2010synthetic} is constructed by normalizing the average post-treatment prediction error by the average pre-treatment one \citep[p.10]{van2024return},
\[
r_j=\frac{R_j\left(T^*+1, T\right)}{R_j\left(1, T^*\right)}
\]
where
\[
R_j\left(t_1, t_2\right)=\left(\frac{1}{t_2-t_1+1} \sum_{t=t_1}^{t_2} d_t^2\right)^{1 / 2}
\]
the root mean squared prediction error in the $2$-Wasserstein distance at time $t$, 
\[
d_t^2=\int_{q_{\min }}^{q_{\max }}\left|\sum_{j=1}^J \lambda_j^* F_{j t}^{-1}(q)-F_{0 t}^{-1}(q)\right|^2 \mathrm{d} q
\]
with $\lambda^*_j$ the optimal DiSCo weights. As explained above, one can test for effects in a restricted part of the distribution by setting \texttt{qmin} $> 0$ or \texttt{qmax} $< 1$. Then we calculate the p-value of the permutation test as
\[
p=\frac{1}{J+1} \sum_{i=0}^J \mathrm{1}\left\{r_j \geq r_0\right\}
\]
which calculates how frequently the normalized placebo effects exceed the true effect on the treated. The null hypothesis that the effect on the treated is drawn from the placebo distribution can be rejected at standard significance levels, for instance $p\leq 0.05$. This p-value is stored in the \texttt{e(pval)} scalar by the \texttt{disco} command when the \texttt{permutation} option is specified. 

\subsubsection{Bootstrapped Confidence Intervals}

The permutation test described above provides an intuitive, finite-sample approach to inference that has become standard in synthetic controls. In addition to the permutation test, we also provide a bootstrap procedure to compute confidence intervals that account for the randomness in the estimation of the distributions. It can be requested by specifying the \texttt{ci} option. Note that this can be computationally intensive, depending on the number of bootstrap replications requested via the \texttt{boots} option. If \texttt{ci} is specified, the \texttt{disco\_plot} and \texttt{disco\_estat} commands will automatically include confidence intervals and/or bootstrapped standard errors.

In \citet[Theorem 1]{van2024return}, we prove that the bootstrap is uniformly valid for the DiSCo method up to a negligible set, meaning that the bootstrapped empirical process,
\begin{equation} \label{eq:bootstrap_gap}
\tilde{\mathbb{G}}_n=\sqrt{n}\left(\tilde{F}_{0 t n, N}^{-1}-\hat{F}_{0 t n, N}^{-1}\right)=\sqrt{n}\left(\sum_{j=1}^J \tilde{\lambda}_{j n}^* \tilde{F}_{j t n}^{-1}-\sum_{j=1}^J \hat{\lambda}_{j n}^* \hat{F}_{j t n}^{-1}\right)
\end{equation}
converges uniformly to the “true” empirical process,
\[
\mathbb{G}_n=\sqrt{n}\left(\hat{F}_{0 t n, N}^{-1}-F_{0 t n}^{-1}\right)=\sqrt{n}\left(\sum_{j=1}^J \hat{\lambda}_{j n}^* \hat{F}_{j t n}^{-1}-\sum_{j=1}^J \lambda_j^* F_{j t n}^{-1}\right)
\]
where $\tilde{\lambda}_{j n}$ and $\tilde{F}_{j t n}^{-1}$ are the bootstrapped optimal weights and quantile functions. 

In practice, the following pseudo-code provides a high-level overview of how the bootstrapped confidence intervals are computed. For a more detailed exposition, see Algorithm 1 in \citet{van2024return}.

\begin{enumerate}
\item Estimate the main DiSCo weights $\lambda^*_j$ on the original pre-treatment data by solving (the empirical version of) \eqref{eq:2-wasserstein} or \eqref{eq:1-wasserstein}.
 \item Construct the main post-treatment estimate — the synthetic quantile function in \eqref{eq:synthetic_quantile}. 
\item For b from 1 to \texttt{boots}:
   \begin{enumerate}
   \item Resample each firm’s data (pre + post) to create a bootstrap sample.
   \item Re-estimate weights on the bootstrapped pre-treatment data.
   \item Construct the estimate for each bootstrapped post-treatment $t>T_0$, using (1) the new weights calculated with the resampled pre-treatment quantile functions and (2) the resampled post-treatment quantile functions:
   $
   \sum_{j=1}^J \tilde{\lambda}_{j} \tilde{F}^{-1}_{jt}.
   $
   \item Compute the bootstrap “gap” in \eqref{eq:bootstrap_gap}.
\end{enumerate}
\item Form confidence intervals/bands using the empirical distribution of the bootstrap “gaps.”
\end{enumerate}

\subsection{Relation to Classical Synthetic Controls}

The DiSCo method reduces to the classical synthetic controls method of \citet{abadie2003economic} when only aggregate data are given — i.e., when each unit $j$ has a single aggregate observation rather than an entire distribution of observations \citep[\S 4.1]{gunsilius2023distributional}. As such, it complements the classical method for settings where researchers have access to sub-aggregate data \textit{within} each unit $j$ and want to (1) use all available data to inform estimation rather than collapsing it down to a single point for each unit and/or (2) estimate distributional treatment effects.

Several synthetic control commands already exist in Stata. The \texttt{synth} command implements the canonical estimator of \citet{abadie2003economic}, which constructs a synthetic control for \textit{scalar-valued} outcomes by constructing a convex weighted average of control units' outcomes, where each unit has a single number as its outcome value in each period. \texttt{gsynth} \citep[now absorbed into the \texttt{fect} package]{xu2017generalized} builds on top of that by explicitly estimating the control units' outcomes through interactive fixed-effects models, which lets it handle several treated units and staggered timing while relaxing the parallel-paths assumption. Nonparametric variants such as \texttt{npsynth} loosen the functional form of the fit, but they too model settings where the outcome is a single number per period for each unit.

\texttt{disco} differs in the object it matches. The unit of estimation is the whole quantile function, or the whole CDF, so the command puts one weight on each control distribution and returns a counterfactual distribution for every post-treatment period instead of a single number. The most common setting where an entire distribution is observed for each unit in each period is the grouped data setting, building on  \citet{hausman1981panel} -- see also \citet{chetverikov2016iv} for grouped quantile IV and \citet{van2025regression} for an analogous setting in the regression discontinuity design. In that sense, the \texttt{disco} command is a direct extension of the \texttt{synth} command's estimation approach to a grouped data setting with distribution-valued outcomes. The two approaches coincide when each unit's distribution is degenerate or one only observes the average: in that case, the objective in \eqref{eq:2-wasserstein} reduces to the squared distance between scalars, and the \texttt{disco} weights are exactly the \texttt{synth} weights when using only the pre-treatment outcomes as predictors. A researcher who only needs a mean effect, or who has one observation per unit, should use \texttt{synth} or \texttt{gsynth}. \texttt{disco} earns its extra cost when each unit holds many observations and the researcher is interested in distributional treatment effects: which part of the tenure distribution loses mass after a return-to-office mandate, for example. 

\subsection{Implementation Details}

The command uses a C++ plugin for the constrained quadratic optimization in \eqref{eq:2-wasserstein} to improve performance, based on the \texttt{quadprog} code by \citet{digaspero2025}. This code implements the Goldfarb–Idani active-set dual method \citep{goldfarb1983numerically}. The C++ plugin is compatible with Linux/Unix, macOS (both Apple Silicon and Intel), and Windows system architectures. 
For the optimization in \eqref{eq:1-wasserstein}, the command relies on the \texttt{LinearProgram} class in Mata. 

Memory requirements scale with the number of units ($J$), time periods ($T$), grid points ($G$), and bootstrap replications ($B$).

Runtime is dominated by the bootstrap. A point estimate is cheap, but each of the \texttt{boots} replications repeats the entire weight estimation, so confidence intervals are where the time goes. On the tenure panel used below, with about one million employee-quarter observations across 32 firms and three quarters, a single quantile estimate runs in about a second; the same specification with \texttt{m(100)} and \texttt{ci boots(300)} takes about ten seconds on a MacBook Pro with an Apple M1~Pro chip and 16~GB of memory. Three options control the trade-off. \texttt{boots()} enters the runtime close to linearly and is the main expense once \texttt{ci} is on. \texttt{m()} fixes how many points approximate the integral in \eqref{eq:integral_simulation}; it defaults to 1000, and lowering it speeds up every weight solve at some cost to the accuracy of that integral. \texttt{g()} sets how finely the quantile functions and CDFs are evaluated. In practice it pays to develop and debug with point estimates on a coarse grid, then run \texttt{ci} with the full \texttt{boots} count once, for the final numbers.

\section{The \texttt{disco} Command}

\subsection{Syntax}

\begin{stsyntax}
\dunderbar{disco} depvar idvar timevar {\it if}\ \optional{\it in}, 
    \underbar{id}target(\varname) 
    \underbar{t}0(\num)
    \optional{{\it options}}
\end{stsyntax}

\hangpara
{\tt depvar} is the outcome variable (numeric).

\hangpara
{\tt idvar} is the unit identifier variable (numeric).

\hangpara
{\tt timevar} is the time-period variable (numeric).

\subsection{Options}

\subsubsection{Required Options}

\hangpara
{\tt idtarget(varname)} specifies the ID value of the treated unit in \textit{idvar}.

\hangpara
{\tt t0(\#)} specifies the first time period of treatment in \textit{timevar}. Note the slight difference in notation with \citet{abadie2003economic}, who use capital $T_0$ to denote the last pre-treatment period.

\subsubsection{Model Options}

\hangpara
{\tt m(\#)} specifies the number of grid points for the approximation of the integral in \eqref{eq:integral_simulation}. See the explanation there. The default is 1000.

\hangpara
{\tt g(\#)} specifies the number of grid points for the evaluation of the quantile functions and CDFs, i.e., for the vectors/functions that are returned by the command. The default is {\tt g(100)}.

\hangpara
{\tt mixture} requests the CDF-based approach using the 1-Wasserstein metric in \eqref{eq:1-wasserstein}
instead of the default quantile-based approach using the 2-Wasserstein metric in \eqref{eq:2-wasserstein}.
This is recommended for categorical outcomes or when distributions are likely mixtures.
The mixture weights are found by solving a linear program with $2m+J$ variables, where $m$ is
set by \texttt{m()}; this becomes slow for large \texttt{m()}, so values up to around \texttt{m(100)}
are recommended with this option.

\hangpara
{\tt nosimplex} removes the constraint that weights must lie in the unit simplex, 
allowing for negative weights while maintaining that the weights sum to one. 
This enables extrapolation beyond the convex hull of control outcomes.

\hangpara
{\tt qmin(\#)} specifies the minimum quantile for the estimation range; 
the default is {\tt qmin(0)}.

\hangpara
{\tt qmax(\#)} specifies the maximum quantile for the estimation range; 
the default is {\tt qmax(1)}.

\subsubsection{Inference Options}

\hangpara
{\tt ci} computes bootstrap confidence intervals for all distributional effects. 
Confidence intervals account for estimation uncertainty in both the weights and empirical distributions.

\hangpara
{\tt boots(\#)} specifies the number of bootstrap replications for confidence intervals. 
The default is {\tt boots(300)}. More replications provide more stable intervals but increase computation time linearly.

\hangpara
{\tt cl(\#)} sets the confidence level as a percentage for intervals. 
The default is {\tt cl(0.95)}. Must be between 0 and 1.

\hangpara
{\tt permutation} performs a permutation test by applying the estimation to each control unit as if it were treated, following \citet{abadie2010synthetic, van2024return}. 
Reports a p-value for the null of no effect.

\hangpara
{\tt seed(\#)} sets the random-number seed for reproducibility of bootstrap inference; the permutation test does not depend on the seed because the integral in \eqref{eq:integral_simulation} is evaluated on a fixed, uniformly spaced grid, so only the bootstrap uses random draws.\footnote{This differs from the companion R package \citep{discosR}, which draws the quantile levels in \eqref{eq:integral_simulation} at random; there, the point estimates and the permutation test do depend on the seed.}

\hangpara
{\tt nouniform} requests pointwise rather than uniform confidence bands. 
Uniform bands control for multiple testing across quantiles but are wider than pointwise bands. Use of this option is discouraged since inference on functional objects should generally be uniform.

\hangpara
{\tt agg} specifies the type of aggregation for summary statistics and plots. One of
\begin{itemize}
\item \texttt{"quantile"}: summarize estimated quantile functions
\item \texttt{"cdf"}: summarize estimated CDFs 
\item \texttt{"quantileDiff"}: summarize differences in quantiles between treated and synthetic
\item \texttt{"cdfDiff"}: summarize differences in CDFs between treated and synthetic
\end{itemize}

\hangpara 
{\tt samples(numlist)} specifies quantile or CDF points to aggregate over for summary statistics. For quantiles, these are in [0,1]. 
For CDFs, these are values of the outcome variable. If not specified, the default is to partition the support 
(either [0,1] or the range of the outcome variable) into 5 equally spaced points ([0, 0.25, 0.5, 0.75, 1]) and aggregate the treatment effects within those intervals. 

\subsection{Stored Results}
\begin{stresults2}
\stresultsgroup{Scalars}\\
\stcmd{e(amin)} &  minimum observed outcome value 
\\ 
\stcmd{e(amax)} & maximum observed outcome value 
\\
\stcmd{e(m)} & number of samples to approximate integrals in estimation step
\\
\stcmd{e(g)} & number of grid points to evaluate quantile function / CDF on
\\
\stcmd{e(t\_max)} & last time period in the dataset
\\
\stcmd{e(N)} & number of observations 
\\
\stcmd{e(pval)} & p-value from permutation test \\
\stcmd{e(doci)} & indicates whether CIs were requested 
\\
\stcmd{e(t0)} & first treatment period \\
\stcmd{e(cl)} & confidence level (e.g., 0.95)
\\
\stresultsgroup{Macros}\\
\stcmd{e(cmd)} & \stcmd{disco} \\
\stcmd{e(agg)} & aggregation type (e.g., "quantile")
\\
\stcmd{e(cmdline)} & command as typed
\\

\stresultsgroup{Matrices}\\
\stcmd{e(cids)} & control IDs (1 $\times$ J)  \\
\stcmd{e(weights)} & synthetic control weights (J $\times$ 1)
\\
\stcmd{e(cdf\_t)} & treated unit CDFs (G $\times$ T) \\
\stcmd{e(cdf\_synth)} & synthetic CDFs (G $\times$ T)
\\
\stcmd{e(quantile\_t)} & treated unit quantiles (G $\times$ T) \\
\stcmd{e(quantile\_synth)} &   synthetic quantiles (G $\times$ T)
\\
\stcmd{e(cdf\_diff)} & CDF differences (G $\times$ T) \\ 
\stcmd{e(quantile\_diff)} & quantile differences (G $\times$ T)
\\
\stcmd{e(summary\_stats)} & summary stats matrix (length of \texttt{samples} option - 1 $\times$ 7)
\\
\end{stresults2}

\subsection{Post-estimation Commands}

Three post-estimation commands are available after \texttt{disco}: \texttt{disco\_estat} for summary tables of the aggregated treatment effects, \texttt{disco\_plot} for visualizations, and \texttt{disco\_weight} for inspecting the estimated weights.

\subsubsection{The \texttt{disco\_estat} command}

The \texttt{disco\_estat} command displays summary statistics after \texttt{disco} estimation.
It provides a detailed summary of aggregated distributional treatment effects, including point estimates, standard errors, and confidence intervals across different time periods. Aggregation is done along the partition of the support specified in the \texttt{samples} option in the \texttt{disco} command.

\paragraph{Syntax}

\begin{stsyntax}
\dunderbar{disco\_estat} \underbar{sum}mary
\end{stsyntax}

\paragraph{Remarks}

\texttt{disco\_estat} requires that \texttt{disco} has been run previously with a specific \texttt{agg} and \texttt{samples} option, which cannot be altered post-estimation. This limitation is imposed to avoid the command having to pass large bootstrap matrices back to Stata when the \texttt{ci} option is specified.

\subsubsection{The \texttt{disco\_plot} command}

The \texttt{disco\_plot} command creates visualizations after \texttt{disco} estimation.
It displays quantile functions, CDFs, and their differences over time, with optional confidence intervals. Plot types are automatically adapted to the aggregation method used (e.g., \texttt{quantileDiff}, \texttt{cdfDiff}), and confidence intervals appear if \texttt{ci} was specified in \texttt{disco}.

\paragraph{Syntax}

\begin{stsyntax}
\dunderbar{disco\_plot} \optional{, {\it options}}
\end{stsyntax}

\paragraph{Options}

The options for \texttt{disco\_plot} are the following.

\hangpara
{\tt agg(string)} specifies if the outcome variable is categorical and a CDF plot is requested in \texttt{agg}. If specified, a bar plot is created instead of a line plot.

\hangpara
{\tt categorical} specifies if the outcome variable is categorical and a CDF plot is requested in \texttt{agg}. 
If specified, a bar plot is created instead of a line plot.

\hangpara
{\tt title(string)} specifies a custom title for the graph. Defaults vary by plot type.

\hangpara
{\tt ytitle(string)} specifies a custom y-axis title. Defaults vary by plot type.

\hangpara
{\tt xtitle(string)} specifies a custom x-axis title. Defaults vary by plot type.

\hangpara
{\tt color1(string)} specifies the color for the first series (default \texttt{"blue"}).

\hangpara
{\tt color2(string)} specifies the color for the second series in level plots (default \texttt{"red"}).

\hangpara
{\tt cicolor(string)} specifies the color for confidence intervals (default \texttt{"gs12"}).

\hangpara
{\tt lwidth(string)} specifies the line width (default \texttt{"medium"}).

\hangpara
{\tt lpattern(string)} specifies the line pattern for the second series (default \texttt{"dash"}).

\hangpara
{\tt legend(string)} specifies legend options, passed directly to Stata’s graph commands; see \rref{legend\_options}.

\hangpara
{\tt byopts(string)} specifies options for small multiples using \texttt{by()}, such as \texttt{rows(2)} or \texttt{ytitle}; see \rref{by\_option}.

\hangpara
{\tt plotregion(string)} specifies options for the plot region in Stata graphs; see \rref{region\_options}.

\hangpara
{\tt graphregion(string)} specifies options for the overall graph region; see \rref{region\_options}.

\hangpara
{\tt scheme(string)} specifies a scheme name for the graph; see \rref{scheme\_option}.

\hangpara
{\tt hline(real)} y coordinate for a dashed grey horizontal line; see \rref{added\_line\_options}.

\hangpara
{\tt vline(real)} x coordinate for a dashed grey vertical line; see \rref{added\_line\_options}.

\hangpara
{\tt xrange(numlist)} numlist of size 2 to set the x range; see \rref{scale\_options}.

\hangpara
{\tt yrange(numlist)} numlist of size 2 to set the y range; see \rref{scale\_options}.

\subsubsection{The \texttt{disco\_weight} command}

The \texttt{disco\_weight} command displays and stores synthetic control weights after \texttt{disco} estimation, matching numeric unit IDs (\textit{id\_var}) with their corresponding string names (\textit{name\_var}). The results are stored in a new frame for further analysis or easy export to tables. The command requires that \texttt{disco} has been run previously.

\paragraph{Syntax}

\begin{stsyntax}
\dunderbar{disco\_weight} \textit{id\_var} \textit{name\_var} \optional{, {\it options}}
\end{stsyntax}

\paragraph{Options}

The options for \texttt{disco\_weight} are the following.

\hangpara
{\tt n({\it \#})} specifies the number of top weights to display and store. The default is 5.

\hangpara
{\tt format(string)} specifies the display format for weights. The default is \texttt{\%12.4f}.

\hangpara
{\tt frame(string)} specifies the name of the frame where results will be stored. The default is \texttt{"disco\_weights"}. If a frame with this name already exists, it will be replaced.

\hangpara
{\tt round({\it \#})} specifies the rounding precision for weights. The default is 0.0001.

\paragraph{Stored results}

\texttt{disco\_weight} stores the following in \texttt{r()}:
\begin{stresults}
\stresultsgroup{Scalars}\\
\stcmd{r(n)} & number of top weights displayed \\
\stresultsgroup{Macros}\\
\stcmd{r(cmd)} & \stcmd{disco\_weight} \\
\stcmd{r(format)} & display format used \\
\stcmd{r(frame)} & name of frame where results are stored \\
\end{stresults}

\section{Empirical Illustration: \citet{van2024return}} \label{sec:application}

\citet{van2024return} study the effect of return-to-office (RTO) mandates on firms' workforce composition. The question that motivates the paper is whether a firm-wide mandate to work from the office at least one day a week causes junior and senior employees to depart the firm at different rates. This is a setting where the treatment (RTO mandate) affects an entire group (a firm) at once, and the outcome of interest is inherently a distribution, in this case the distribution of employee tenure and titles. The paper estimates a distributional synthetic control for three large tech firms that were early to introduce firm-wide RTO mandates, Microsoft, SpaceX, and Apple. The paper's main finding is that the RTO mandates at these large technology firms were followed by a significant decrease in employee tenure at the top tail of the distribution, with employees at the upper deciles departing Microsoft approximately 2 months earlier due to the RTO mandate; and a shift towards more junior titles, with an increase in employees at or below the manager level of around 4\%. We reproduce those results below.

The do-files and (perturbed) data that reproduce this section are distributed with this article.

\subsection{Data and Methods} 
\paragraph{Data Overview.} In the paper, we use quarterly résumé data from People Data Labs,\footnote{\url{https://www.peopledatalabs.com/}} a large provider of individual-level employment records that covers over half of the total headcount for most major tech firms (see \citealt{van2024return} for details). Each firm-quarter includes a repeated cross-section of employees, from which we observe job tenure and a seniority rank (ranging from “unpaid” to “CXO”). We merge in information on return-to-office (RTO) mandates gathered from public announcements, employee forums, and the Flex Index by Scoop Technologies, as well as layoff records from \url{layoffs.fyi}. Our main estimation sample focuses on the return to office at Microsoft in April 2022, plus a control pool of firms never adopting RTO within the observation window. For more details on the data, see \citet{van2024return}. In practice, we use a perturbed version of the original data for confidentiality reasons, where we add uniform noise to the outcome variables. This preserves the shape of the distributions while keeping the dataset shareable.

\paragraph{Empirical Approach.} Using these data, we then apply a distributional synthetic controls design that computes synthetic control weights to closely match Microsoft’s distribution of employee tenure in the two quarters before its RTO. The post-RTO gap between the observed and synthetic distributions isolates the policy’s causal effect on workforce composition. 

In the notation from Section \ref{sec:methodology}, $i$ refers to a single employee at firm $j$ (with $j=1$ corresponding to Microsoft), $t$ refers to a quarter of the year, and $Y_{ijt}$ is one of two main outcomes:
\begin{enumerate}
    \item $Tenure_{ijt}$, the number of days employee $i$ has been employed at company $j$.
    \item $Title_{ijt}$, an integer classification of employee $i$'s job title at firm $j$, ranging from 1 (trainee) to 10 (C-suite). See \citet[Table 1]{van2024return} for the full classification.
\end{enumerate}
We solve \eqref{eq:2-wasserstein} for $Tenure_{ijt}$ (as it is continuous) and \eqref{eq:1-wasserstein} for the categorical $Title_{ijt}$. Put differently, it makes sense to interpolate over the support of tenure (e.g., 5 days, 5.5 days), but it generally does not make sense to talk about an employee being “25\% junior associate and 75\% manager.”

\subsection{Results: Quantile Approach}

First, we reproduce the main Figure 4 in the paper. To that end, we start by loading and inspecting the (perturbed) data:
\begin{stlog}
. use "tenure_anonymized.dta", clear
{\smallskip}
. list in 1/5, ab(20)
{\smallskip}
     {\TLC}\HLI{44}{\TRC}
     {\VBAR} time_col   id_col   company_name     y_col {\VBAR}
     {\LFTT}\HLI{44}{\RGTT}
  1. {\VBAR}        2       17         oracle   1682.25 {\VBAR}
  2. {\VBAR}        3        1       deloitte   375.783 {\VBAR}
  3. {\VBAR}        2       72             3m   5276.48 {\VBAR}
  4. {\VBAR}        3        2      microsoft    957.55 {\VBAR}
  5. {\VBAR}        2        2      microsoft   2745.96 {\VBAR}
     {\BLC}\HLI{44}{\BRC}
\end{stlog}

The data structure shows the essential columns required for estimation: (1) a \textit{time column}, indicating the time period (quarters of 2022); (2) a numeric ID column indicating the ID of the aggregate unit (firms, e.g. ID ``17'' for Oracle); (3) a numeric outcome variable (days the employee at Oracle had been working there at the start of quarter 2). There is also an optional string column indicating the name of the aggregate unit (“Oracle”), which can be used in the \texttt{disco\_weight} command to inspect synthetic control weights by name.

With these data, we can replicate the figure with:
\begin{stlog}
. disco y_col id_col time_col, idtarget(2) t0(3) agg("quantileDiff") ///
>         seed(12143) g(10) m(100) ci boots(300)
\end{stlog}
Here, we evaluate the quantile functions on 10 grid points using \texttt{g(10)}, but use 100 points for estimation (\texttt{m(100)}) to approximate the integral in \eqref{eq:integral_simulation}. In the paper, we used 1,000 samples; we lower that to 100 here for replication speed. Similarly, we compute confidence intervals via the \texttt{ci} option with only 300 bootstrap repetitions (\texttt{boots(300)}) instead of 1,000 as in the paper.

After estimation, we first examine the top 5 weights by company name:

\begin{stlog}
. disco_weight id_col company_name, n(5)
{\smallskip}
Top 5 weights:
\HLI{50}
{\smallskip}
  {\TLC}\HLI{19}{\TOPT}\HLI{8}{\TRC}
  {\VBAR}              name {\VBAR} weight {\VBAR}
  {\LFTT}\HLI{19}{\PLUS}\HLI{8}{\RGTT}
  {\VBAR}            amazon {\VBAR}  .2203 {\VBAR}
  {\VBAR}          autodesk {\VBAR}  .1271 {\VBAR}
  {\VBAR}             cisco {\VBAR}  .1066 {\VBAR}
  {\VBAR} dell technologies {\VBAR}  .0991 {\VBAR}
  {\VBAR} slalom consulting {\VBAR}  .0962 {\VBAR}
  {\BLC}\HLI{19}{\BOTT}\HLI{8}{\BRC}
\end{stlog}

These weights are very close to those in \citet[Table 3]{van2024return}, despite the data perturbation. As noted there, firms like Amazon, Cisco, and Dell receive high weights because they are highly similar to Microsoft (the treated firm).

We then directly plot the quantile effects, defined as the differences between the treated and synthetic quantile functions at each of the 10 deciles:

\begin{stlog}
. disco_plot, title(" ") ytitle("Difference in Tenure (Days)") ///
>      hline(0) scheme("stsj")
\end{stlog}

\begin{figure}[ht!]
    \centering
    \includegraphics[width=0.99\linewidth]{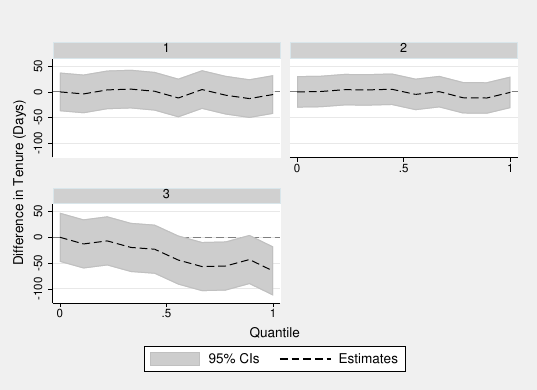}
    \caption{Replication of Tenure Result}
    \label{fig:tenure_quantileDiff}
\end{figure}

This command creates Figure \ref{fig:tenure_quantileDiff}, which closely resembles Figure 4 in the paper.

As discussed in the paper, panel 3 in the figure suggests a statistically significant decrease in tenure (about 50 days) for employees in the top 2--3 deciles of the tenure distribution, but no such decrease at the lower deciles. In other words, the RTO caused longer-tenured employees to leave Microsoft, but not those with shorter tenures. Because these effects are calculated as the difference between the observed and the synthetic quantile function, they have a counterfactual interpretation — i.e., the drop in tenure at the top deciles is relative to a “synthetic” Microsoft that did not return to the office. Finally, panels 1 and 2 in the figure show the quantile differences for the two quarters before the RTO. Since 0 is included in the 95\% confidence bands there, we see that the synthetic control replicated Microsoft’s tenure distribution well, supporting its use for constructing post-treatment counterfactuals.

\begin{figure}[ht!]
    \centering
    \includegraphics[width=0.99\linewidth]{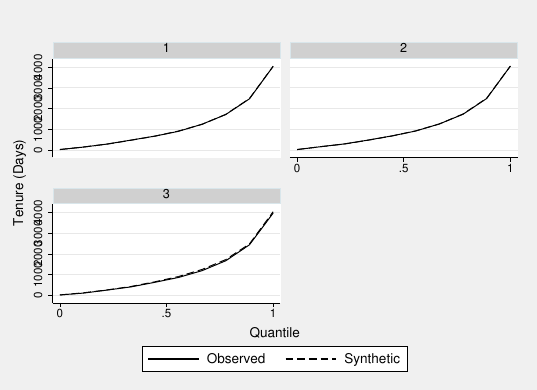}
    \caption{Replication of Tenure Result}
    \label{fig:tenure_quantile}
\end{figure}

Although not shown in the paper, a researcher may be interested in comparing the synthetic and observed quantile functions side by side. This can be done with:
\begin{stlog}
. disco_plot, title(" ") ytitle("Tenure (Days)") agg("quantile")
\end{stlog}
which yields Figure \ref{fig:tenure_quantile}.

We observe similar patterns here: a strong pre-treatment fit, with the two quantile functions overlapping, and a downward shift in the synthetic quantile function at the top deciles after the RTO. Additionally, it clarifies the absolute magnitudes in the distribution: the top 3 deciles, where the largest effects occur, correspond to employees who have been at Microsoft for at least 1,500 days (about 4 years).

Finally, a researcher may want to study the treatment effects in aggregate parts of the distribution. To do so, we can run:

\begin{stlog}
. disco_estat summary
{\smallskip}
Summary of quantile effects
\HLI{72}
   Period  Range              Effect   Std. Err.   [95\% Conf. Interval]
\HLI{72}
        3  0.00-0.25          -6.659      24.256     -54.201     40.883
        3  0.25-0.50         -21.443      24.256     -68.985     26.099
        3  0.50-0.75         -50.509      24.256     -98.052     -2.967*
        3  0.75-1.00         -54.751      24.256    -102.293     -7.209*
\HLI{72}
* denotes significance at the 95\% confidence level
\end{stlog}

By default, the command aggregates the treatment effects into four buckets corresponding to the four quartiles. We see statistically significant effects in the top two quartiles, corresponding to decreases of about 51 and 55 days in employee tenure; the effect in the third quartile is borderline, with the confidence interval barely excluding zero.

\subsection{Results: CDF Approach}

An analogous exercise can be carried out for the categorical $Title$ variable, except that we now want the command to solve \eqref{eq:1-wasserstein}, as this variable takes integer values from 1 to 10. Hence, we specify the \texttt{mixture} option and force the command to use a grid of 10 points by setting \texttt{g(10)} and \texttt{m(10)}. Concretely:

\begin{stlog}
. use "titles_anonymized.dta", clear 
{\smallskip}
. list in 1/5, ab(20) 
{\smallskip}
     {\TLC}\HLI{42}{\TRC}
     {\VBAR} time_col   id_col   company_name   y_col {\VBAR}
     {\LFTT}\HLI{42}{\RGTT}
  1. {\VBAR}        3       17         oracle       8 {\VBAR}
  2. {\VBAR}        2        1       deloitte       5 {\VBAR}
  3. {\VBAR}        3        1       deloitte       6 {\VBAR}
  4. {\VBAR}        1        1       deloitte       5 {\VBAR}
  5. {\VBAR}        2      245         splunk       4 {\VBAR}
     {\BLC}\HLI{42}{\BRC}
{\smallskip}
. 
. disco y_col id_col time_col, idtarget(2) t0(3) agg("cdfDiff") seed(12143) /// 
>         mixture g(10) m(10) ci boots(300) 
. disco_plot, title(" ") ytitle("Change in CDF") hline(0) categorical ///
>         scheme("stsj") color("bluishgray") 
\end{stlog}

\begin{figure}[ht!]
    \centering
    \includegraphics[width=0.99\linewidth]{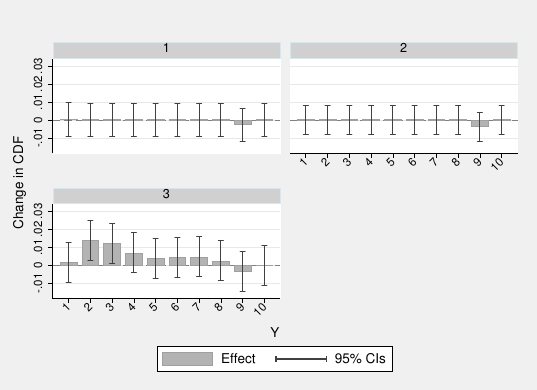}
    \caption{Replication of Titles Result}
    \label{fig:title_cdfDiff}
\end{figure}

Since \texttt{y\_col} is clearly categorical, specifying \texttt{categorical} in the \texttt{disco\_plot} command generates a bar plot rather than a line plot. The resulting figure, shown in Figure \ref{fig:title_cdfDiff}, closely resembles Figure 5 in \citet{van2024return}, though the data perturbation slightly widens the confidence intervals.

This figure can be interpreted as showing a statistically significant increase in the mass under the CDF for employees with titles below the (senior) manager level (4), but no such increase for higher-ranked employees. By the nature of cumulative distribution functions, this means Microsoft’s workforce rebalanced toward lower-ranked employees after the RTO, likely due to an outflow of senior talent.

\section{Conclusions}

We have presented \texttt{disco}, a Stata implementation of the Distributional Synthetic Controls estimator of \citet{gunsilius2023distributional}, and show how it can be used to reproduce the distributional effects of RTO mandates in \citet{van2024return}.

Two extensions would make the command more useful. The first is a multivariate version that matches the joint distribution of several outcomes at once instead of one at a time, following \citet{gunsilius2024tangential}. That turns the weight problem into a multidimensional optimal-transport problem, which is harder to solve and will need a dedicated routine. The second is faster inference. The bootstrap still drives the runtime: the command already sorts and caches the outcome data across replications, but spreading the bootstrap loop over multiple cores would cut the remaining time by a further large factor.

\section{Acknowledgments}

This research was supported in part through computational resources and services provided by Advanced Research Computing at the University of Michigan, Ann Arbor.

\bibliographystyle{sj}
\bibliography{sj}

\begin{aboutauthors}
Florian Gunsilius is an Associate Professor in the Department of Economics at Emory University. His research interests are nonparametric approaches for statistical identification, estimation, and inference. His current focus is on statistical optimal transport theory, mean field estimation, causal inference, and free discontinuity problems.

David Van Dijcke is an Assistant Professor in the Department of Economics at the University of Virginia. His research interests focus on causal inference with non-Euclidean and distributional data, nonparametric estimation, and regression discontinuity designs.
\end{aboutauthors}

\clearpage
\end{document}